\documentclass[apj]{emulateapj}
\usepackage{booktabs}

\shorttitle{Parameter Estimation in the MNL Regime}
\shortauthors{ S.~ Safi, M.~ Farhang}
\usepackage{graphicx}
\usepackage{grffile}

\usepackage{color}
\usepackage[colorlinks=true,
            linkcolor=black,
            urlcolor=black,
            citecolor=blue, breaklinks=true]{hyperref}
\usepackage{url}
\usepackage{amsmath}
\usepackage{amssymb}
\usepackage{gensymb}

\usepackage{graphics}
\usepackage{graphicx}
\usepackage{natbib}
\usepackage{psfig}
\input{epsf}
\usepackage{ae,aecompl}
\usepackage{booktabs}

\newcommand{\be}{\begin{equation}}
\newcommand{\ee}{\end{equation}}

\begin{document}

\title{Sensitivity of cosmological parameter estimation to  nonlinear prescription \\
from Galaxy Clustering}

\author{Sarah Safi, Marzieh Farhang}
\affil{\scriptsize
{Department of Physics, Shahid Beheshti University, 1983969411,  Tehran Iran}\\
}

\email{Corresponding author email: m\_farhang@sbu.ac.ir}

\begin{abstract}
Next generation large scale surveys will
probe the nonlinear regime with high resolution. 
Making viable cosmological inferences based on these observations requires accurate theoretical modeling of the mildly nonlinear regime.
In this work we investigate the sensitivity of cosmological parameter measurements from future probes of galaxy clustering to the choice of nonlinear prescription up to $k_{\rm max}=0.3~h~\rm{Mpc}^{-1}$. In particular, we calculate the induced parameter bias when the mildly nonlinear regime is modeled by the Halofit fitting scheme. We find significant ($\sim5\sigma$) bias for some parameters with a future Euclid-like survey. 
We also explore the contribution of different scales to the parameter estimation for different observational setups and cosmological scenarios, compared for the two nonlinear prescriptions of Halofit and EFTofLSS. 
We include in the analysis the free parameters of the nonlinear theory and a blind parametrization for the galaxy bias. 
We find that marginalization over these nuisance parameters significantly boosts the errors of the standard cosmological parameters. This renders the differences in the predictions of the various nonlinear prescriptions  less effective when transferred to the parameter space. More accurate modeling of these nuisance parameters would therefore greatly enhance the cosmological gain from the mildly nonlinear regime.
 \end{abstract}

\keywords{mildly nonlinear regime - cosmology; theory - large scale structure }

\section{Introduction}
Future large scale  surveys, such as Euclid  \citep{redbook} and the square kilometer array (SKA) \citep{santos15}, will probe the mildly nonlinear (MNL afterwards) regime of structure formation with unprecedented accuracy.
This high accuracy would translate into tight constraints on cosmological parameters.
 A  large amount of study has gone into exploring this regime for making cosmological inferences in different frameworks \citep[see, e.g., ][among many others] {martinelli11,majerotto12,di12,wang12,
dePutter13,ska16,sartoris16,eftde16,blanchard19,casas17,
sprenger19}.

On the other hand, the analysis of data from these high--resolution experiments calls for  accurate theoretical modeling that reliably describes the MNL regime for optimal extraction of the available cosmological information. 
This regime has been investigated perturbatively by different approaches including the  SPT \citep{bernardeau2002,carlson09}, the RPT \citep{renormalized06}, the RegPT\citep{regpt12} and the EFTofLSS \citep{eft12}.
For a list of theoretical methods and the scales of the validity of their predictions see \cite{carlson09}. 

The effective field theory of large scale structure, abbreviated as the EFTofLSS, is the main nonlinear analytical framework used in this work. It provides a convergent perturbation theory and is claimed to be more successful compared to various  alternatives \citep{baumann10,eft12,twoloop14}. 
This theory generalizes the SPT by effectively taking into account the impact of short modes on large scales through the introduction of certain free coefficients to be measured by data
\citep{eft12,twoloop14}. 
The EFTofLSS is so far developed at the two-- and  three--loop order  \citep{twoloop14,baldauf15effective,threeloop19} with various modifications such as bias and baryonic effects taken  into account \citep{senatorebias14,mirbabayi14,assassi14,angulo15bias,lewandowski15}.
Various predictions of the theory have also been compared to simulations, including the dark matter density power spectrum \citep{eft12,senatore14via,twoloop14}, bispectrum \citep{angulo14,baldauf14}, dark matter momentum power spectrum \citep{senatore14via} and the dark matter power spectrum in redshift space \citep{senatore14vja}.
Recently, the EFTofLSS has been applied to the analysis of the BAO data in order to examine the extent to which this theory can provide a theoretical template in cosmological parameter estimation \citep{boss19d'amico,boss19colas,blinded20,limits20,hubble20}.

In this work we study the impact of the prescription for the MNL regime on cosmological parameter estimation from the galaxy clustering probe of  Euclid-- and SKA--like surveys. 
(In a recent work, \cite{euclidwl20} studied a similar problem but for the Euclid weak lensing probe and with different nonlinear prescriptions.)
We investigate how the inaccuracies in the nonlinear prescription lead to biases in the parameters. 
We also explore the available information in these regimes to constrain parameters in a couple of different cosmological scenarios.  
The impact of marginalization over parameters describing the galaxy bias and the free parameters of the nonlinear theory is also considered.  

The rest of this paper is organized as follows.  In Section~\ref{sec:bias} we investigate the induced bias in the parameter estimations due to improper choice of nonlinear prescription and in Section~\ref{sec:k-dep} we explore the cosmological information available for parameter measurements in the MNL regime, under different assumptions for  experimental specifications and cosmological and analysis framework. We conclude in Section~\ref{sec:conclude}. 
\begin{figure*}
   \begin{center}
\includegraphics[width=\textwidth]{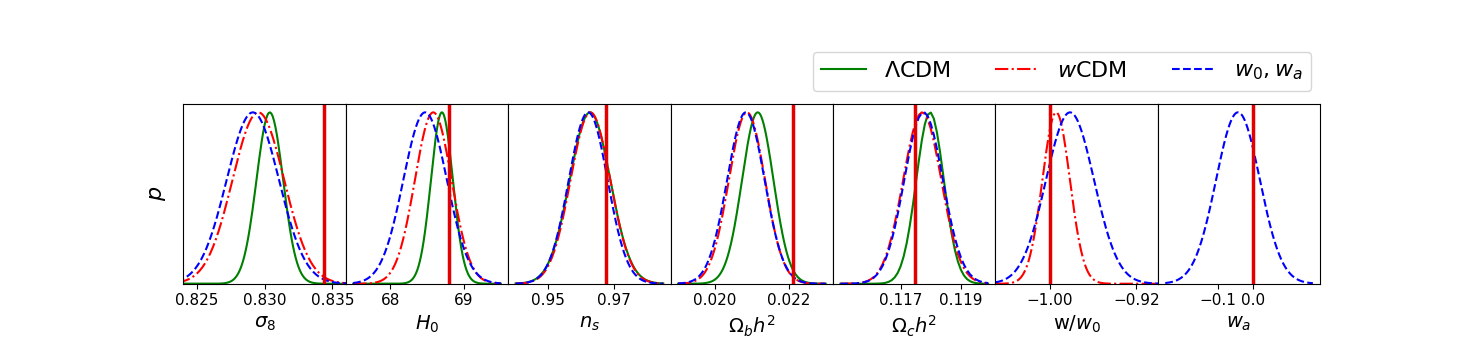} 
    \caption{1D marginalized  parameter likelihoods for three different cosmological scenarios with $k_{\rm{max}}=0.3~h~\rm{Mpc}^{-1}$. Vertical lines illustrate the parameter fiducial values used in the simulation. This analysis does not include any free nuisance parameters. }
    \label{fig:likp3}
    \end{center}
\end{figure*}
\section{nonlinear prescription bias in parameter estimation}\label{sec:bias}
The accuracy and robustness of the nonlinear recipe of  structure formation   
is of particular importance in 
making forecasts and analyzing the high-resolution data of future surveys.
Here we investigate the insufficiencies of a particular prescription, the Halofit model (HF afterwards),   
in describing its impact on  cosmological inferences.
HF is a commonly used recipe for modeling the matter power spectrum in the MNL regime \citep{smith03,takahashi12}  and is based on fitting formulas from the results of $N$-body simulations \citep{ma2000,seljak2000,cooray02}.

We quantify this effect by 
calculating the induced shifts in the parameter estimation for several different cosmological models.  
As the dataset we use power spectrum simulations of galaxy clustering for a future Euclid--like survey. 
We review the assumptions used to simulate the observed galaxy power spectrum in Section~\ref{sec:sim}
and the cosmological scenarios used for the analysis of the simulations in Section~\ref{sec:DEmodels}.  The results are presented and discussed in Section~\ref{sec:res1}. 

\subsection{Simulations of Galaxy Clustering}\label{sec:sim}
Here we use the HF precription to make predictions for a future Euclid-like large scale survey.
We  simulate our mock data based on the theoretical matter power spectrum from the ds14-a set of Dark Sky $N$-body simulation series \citep{darksky}. 
The fiducial values of the cosmological parameters used in the simulations are $\{\Omega_m,\Omega_b,\Omega_\Lambda, h, n_s,\sigma _8\} = \{0.295, 0.0468, 0.705, 0.688, 0.9676, 0.8344\}$.
From these power spectra, we simulate the observed power spectra of galaxy clustering (or GC) for Euclid--like specifications (the first column of Table~\ref{table:spec}).

The observed galaxy power spectrum, $P_{\rm {obs}}$, is given by \citep{seo07}
\begin{eqnarray}
P_{\rm {obs}}(k,\mu,z)&=&\frac{D_{A,\rm ref}^2 H(z)}{D_{A}^2 H_{\rm ref}(z)}b^2(z)(1+\beta(z) \mu^2)^2 \nonumber\\ 
&\times& e^{-k^2\mu^2(\sigma _{\rm r}^2+\sigma _{\rm v}^2)}P(k,z) \label{eq:pobs}
\end{eqnarray}
where $\mu$ is the cosine of the angle between the line of sight and the wavevector $\vec{k}$, $H(z)$ is the Hubble parameter at redshift $z$ and $D_A(z)$ represents the angular diameter distance to redshift $z$.  We also have $\beta (z)={f(z)}/{b(z)}$ where $f(z)$ represents the linear growth rate and $b(z)$ is the redshift--dependent galaxy bias.
In this section we assume a known $z$-dependence for the galaxy bias, $b(z)=\sqrt{1+z}$ \citep{amendola18}. 
The coefficient ${D_{A,\rm ref}^2 H(z)}/{D_{A}^2 H_{\rm ref}(z)}$ encodes the geometrical distortions due to the Alcock-Paczynski effect \citep{alcock79}.
The effect of redshift measurement uncertainty, $\sigma_z$, propagates into the comoving distance error  $\sigma _r=\frac{c}{H(z)}\sigma_z$ \citep{wang13}.
For the peculiar velocity dispersion we take $\sigma_v =300~{\rm km}/s$ \citep{casas17}. 
Table~\ref{table:spec} presents the experimental specifications used for the GC probe in this work.
\begin{table}[h]
\centering
\begin{tabular}{c c c c }  
\hline
 & \textbf{Euclid} & \textbf{SKA1} &\textbf{SKA2} \\
\hline
$f_{\rm sky}$ & 0.36 &0.12 &0.73 \\
$\sigma _z$ & $0.001(1+z)$ & $0.001(1+z)$& $0.001(1+z)$  \\
$\{ z_{\rm min},z_{\rm max}\}$ & $\{0.7,2\}$ & $\{0,0.7\}$ &$\{0.1,2\}$ \\
$\Delta z$ & $0.1$  &0.1 &0.1\\
\hline  
\end{tabular}
\caption{Euclid, SKA1 and SKA2 specifications for galaxy clustering where $f_{\rm sky}$ is the sky fraction, $\sigma _z$ stands for the spectroscopic redshift error, $\{ z_{\rm min},z_{\rm max}\}$ represents the observed redshift interval and $\Delta z$ is the width of the redshift bins \citep{amendola18,ska16}.}
\label{table:spec}
\end{table}
\subsection{Models of Dark Energy}\label{sec:DEmodels}
In this work we consider three sets of cosmological scenarios, $\Lambda$CDM, $w$CDM with $w\neq -1$ a constant, and a dynamical dark energy model with its equation of  state modeled as \citep{cpl01,lindercpl}
\begin{equation}
w(a)=w_0+(1-a)w_a.
\end{equation}
Here $w_0$ and $w_a$ are constants and the deviation of $(w_0,w_a)$ from $(-1,0)$ encodes the departure from $\Lambda$CDM.
\subsection{Analysis and Results}\label{sec:res1}
The goal in this section is to calculate how the improper choice of the MNL prescription impacts the parameter estimation. Specifically, we calculate the induced bias  in the parameter measurements compared to their fiducial values.
For this purpose, we base our analysis on the publicly available code CosmoMC\footnote{\url{https://cosmologist.info/cosmomc/}}\citep{lewis02,lewis13} to sample the parameter space. CosmoMC uses HF to model the nonlinear regime.  
We map the parameter distribution measured by the GC power spectrum constructed from the Dark Sky simulations with Euclid specifications (Eq.~\ref{eq:pobs}). 
The analysis is performed at four redshift bins centered at $\{0.66,1,1.5,2\}$ and with $k_{\rm max}=0.3~h~\rm{Mpc}^{-1}$,
for the three cosmological models of $\Lambda$CDM, $w$CDM and $(w_0,w_a)$CDM. 

The results of the parameter estimation are illustrated in Figure~\ref{fig:likp3}. The vertical lines indicate the fiducial parameter values used in the simulations.
The measurements are found to be biased with respect to the fiducial values with different levels of significance.
The  few percent mismatch in the power spectra in the MNL regime up to $k_{\rm max}=0.3~h~\rm{Mpc}^{-1}$ between the predictions of HF and $N$-body simulations (assumed to represent data here) leads to a few $\sigma$ bias in the parameter values for the GC probe of a Euclid-like survey. The two parameters $\sigma_8$ and $\Omega_{\rm b}h^2$ suffer from the largest biases, with $\sim (4-5)\sigma$ significance.
This clearly demonstrates the need for very accurate modeling of this cosmologically interesting regime. 

In the next section we investigate the impact of various scales in the MNL regime on parameter estimation for Euclid-- and SKA--like surveys. 
\section{scale-dependence of parameter measurements}\label{sec:k-dep}
Here we explore the information available at various scales for constraining the cosmological parameters. 
As the main theoretical MNL prescription, we use the EFTofLSS and compare its predictions to those from HF.
\subsection{The Nonlinear Prescription}
A sound analysis of  data from a Euclid-like experiment requires high, $\sim 1\%$ precision  modeling of the matter power spectrum.  The accuracy of the popular HF model is claimed to be $\sim 3\%$, rendering it unreliable for the analysis of the next generation high resolution large scale surveys.
Moreover,  HF is calibrated for $\Lambda \rm CDM$  and lacks suitable extensions for non-$\Lambda$CDM models \citep{amendola18}. 
Among the analytical approaches to investigate the MNL regime is the EFTofLSS, abbreviated as EFT in the rest of this section, which  provides an effective description of the universe at large scales through integrating out short-wavelenghth perturbations.
Long-wavelengths are interpreted as an effective fluid characterized by few parameters such as the equation of state, the speed of sound and viscosity.
The coefficients describing the short distance dynamics need to be fitted to observations or measured through $N$-body simulations \citep{baumann10,eft12}.
The EFT predictions are claimed to have considerable improvement over the predictions from alternative models such as SPT \citep{eft12,twoloop14,foreman16precision}. 

In the rest of this work we use EFT as the main prescription to probe into the MNL regime, along with HF for comparison. 
In this work we follow the  parametrization of \cite{foreman16precision} to effectively encapsulate the short-wavelength modes. We refer to these parameters as sound speeds and label them generically by $c_{\rm s}$ throughout the paper.
 The $c_{\rm s}$ represent the three parameters, $c_{\rm s(1)}^2$, $c_1$ and $c_4$,  that characterize the one and two--loop counterterms in the nonlinear corrections to the power spectrum \citep[see Eq.~6.1 of ][]{foreman16precision}. 
 At one--loop, $c_{\rm s(1)}^2$ is the lowest order counterterm while $c_1$ and $c_4$ are the additional higher derivative counterterms appearing at two--loop calculations. The redshift dependence of the counterterms is parametrized with fitting functions with 12 free parameters \citep[see Eq.~5.7 of][]{foreman16precision}.
 \subsection{Analysis}
 Our goal in this section is to determine the $k$-dependence of the estimated errors for various parameters as measured by the GC probe of  Euclid- and SKA-like surveys.
 For this we use the Fisher matrix formalism to estimate the correlation matrix and thus the errors of the parameters. 
 If the distribution of the parameters can be approximated by a Gaussian, their covariance matrix 
$\boldsymbol{C}$ is approximated by the inverse of the Fisher matrix $\boldsymbol{C}={\boldsymbol F}^{-1}$.
In particular, the diagonal elements of the Fisher inverse correspond to the predicted errors for each parameter, 
$\sigma_i^2=(F^{-1})_{ii}$. 
The Fisher matrix of the cosmological parameters ${\boldsymbol q}=\{q_1, ..., q_N\}$ for the GC at redshift $z$  can be written as  \citep{amendola12,seo07}
\begin{equation}
\begin{aligned}
F_{ij}(z)=\frac{V_{\rm sur}}{8\pi^2}\int_{-1}^{+1}d\mu\int_{k_{\rm min}}^{k_{\rm max}}dkk^2\frac{\partial \ln P_{\rm obs}(k,\mu,z)}{\partial q_i}\\
\frac{\partial \ln P_{\rm obs}(k,\mu,z)}{\partial q_j}
\times \left[ \frac{n(z)P_{\rm obs}(k,\mu,z)}{n(z)P_{\rm obs}(k,\mu,z)+1}\right]^2
\end{aligned}
\end{equation}
where $V_{\rm sur}$ is the volume of sky covered by the galaxy survey in bins of width $\Delta z$  centered at $z$. 
The galaxy number density $n(z)$ is taken from Table $3$ of \cite{amendola18} for the Euclid-like case and Tables $2$ and $3$ of \cite{ska16} for the SKA-like cases. 
 For the largest scales used here we take $k_{\rm min}=0.0079~h~\rm Mpc^{-1}$. 
 We consider two scenarios, linear with $k_{\rm max}=0.15~h~\rm Mpc^{-1}$ and MNL with $k_{\rm max}=0.3~h~\rm Mpc^{-1}$. 
The full Fisher matrix includes contributions from all redshift bins of the survey, $\boldsymbol{F}=\sum \boldsymbol{F}(z)$, with $14$ and $26$ bins for the Euclid and SKA-like cases respectively. See Table~\ref{table:spec} for more experimental specifications used in this work.
The fiducial values in our Fisher forecast are consistent with the {\it Planck} measurements \citep{planck18}. 


The galaxy bias is assumed to have a known redshift dependence, similar to Section~\ref{sec:sim} for the Euclid-like case and as Tables $2$ and $3$ of \cite{ska16} for the SKA-like cases.
In parallel, we also investigate the impact of uncertainties in modeling the bias on cosmological measurements.  
For this purpose, in the HF--based analysis we compare the results with a second treatment of the galaxy bias where no prior assumption is imposed on the functional form of $b(z)$.
Instead, we assume $b(z)$ to have a free constant value in each observation bin, and  simultaneously measure it  with the other cosmological parameters. 
The final results from this treatment are marginalized over these bias parameters. 
For the case of EFT analysis, we also follow two paths of treating the $12$  sound speed parameters of EFT, i.e., the $c_{\rm s}$, as fixed and free parameters to be marginalized over. 
%
\begin{table}
\centering
\setlength\tabcolsep{2pt}
\begin{tabular}{l l l l l l l}
\multicolumn{7}{c}{$\Lambda$CDM}\\ \hline \hline
 && {$n_s$}  & $H_0$ & $\Omega _bh^2$& $\Omega _ch^2$ & $\sigma_8$\\
\hline
\textbf{Euclid}\\
\hline
linear & EFT+$c_{\rm s}$ & $1.6\%$ & $0.5\%$ &$3.0\%$&$0.7\%$&$1.2\%$
\\& EFT+bias & $0.8\%$ & $0.5\%$ &$1.8\%$&$1.2\%$&$2.1\%$\\
& EFT & $0.5\%$ & $0.3\%$ &$1.5\%$&$0.5\%$&$0.6\%$\\& HF+bias & $0.7\%$ & $0.5\%$ &$1.8\%$&$1.2\%$&$2.1\%$\\& HF & $0.5\%$ & $0.3\%$ &$1.4\%$&$0.5\%$&$0.6\%$\\
\hline
MNL  & EFT+$c_{\rm s}$ & $0.3\%$ & $0.1\%$ &$0.9\%$&$0.2\%$&$0.4\%$
\\& EFT+bias & $0.3\%$ & $0.1\%$ &$0.8\%$&$0.2\%$&$0.6\%$\\
& EFT&$0.2\%$&$0.1\%$&$0.8\%$&$0.2\%$&$0.3\%$\\& HF+bias & $0.2\%$ & $0.1\%$ &$0.6\%$&$0.2\%$&$0.6\%$\\
& HF&$0.2\%$&$0.1\%$&$0.6\%$&$0.2\%$&$0.3\%$\\
\hline
\textbf{SKA1}\\
\hline
linear  & EFT+$c_{\rm s}$ & $15.5\%$ & $3.8\%$ &$22.4\%$&$8.4\%$&$17.7\%$
\\& EFT+bias & $8.5\%$ & $3.9\%$ &$15.7\%$&$11.2\%$&$21.5\%$\\
& EFT & $5.3\%$&$3.0\%$&$14.4\%$&$5.1\%$&$7.5\%$\\& HF+bias &  $7.7\%$ & $3.7\%$ &$13.6\%$&$10.5\%$&$20.1\%$\\& HF&$4.8\%$&$2.5\%$&$12.0\%$&$4.7\%$&$6.5\%$\\
\hline
MNL & EFT+$c_{\rm s}$ & $4.0\%$ & $0.7\%$ &$11.0\%$&$3.6\%$&$6.2\%$
\\& EFT+bias & $3.1\%$ & $0.7\%$ &$9.7\%$&$3.9\%$&$8.7\%$\\
& EFT&$3.0\%$&$0.6\%$&$9.6\%$&$3.1\%$&$5.4\%$\\& HF+bias & $3.6\%$ & $0.6\%$ &$7.0\%$&$4.5\%$&$9.9\%$\\& HF&$2.7\%$&$0.6\%$&$7.0\%$&$2.8\%$&$4.4\%$\\
\hline
\textbf{SKA2}\\
\hline
linear  & EFT+$c_{\rm s}$ & $1.4\%$ & $0.5\%$ &$2.8\%$&$0.8\%$&$1.4\%$
\\& EFT+bias & $0.9\%$ & $0.5\%$ &$1.8\%$&$1.2\%$&$2.3\%$\\
& EFT & $0.6\%$&$0.4\%$&$1.6\%$&$0.6\%$&$0.8\%$ \\& HF+bias & $0.7\%$ & $0.4\%$ &$1.4\%$&$1.0\%$&$1.8\%$\\& HF&$0.4\%$&$0.3\%$&$1.1\%$&$0.4\%$&$0.5\%$\\
\hline
MNL & EFT+$c_{\rm s}$ & $0.4\%$ & $0.1\%$ &$1.0\%$&$0.3\%$&$0.5\%$
\\& EFT+bias & $0.3\%$ & $0.1\%$ &$0.9\%$&$0.3\%$&$0.8\%$\\
& EFT&$0.3\%$&$0.1\%$&$1.0\%$&$0.2\%$&$0.5\%$\\& HF+bias & $0.2\%$ & $0.1\%$ &$0.6\%$&$0.2\%$&$0.6\%$\\& HF&$0.2\%$&$0.2\%$&$0.1\%$&$0.2\%$&$0.3\%$\\
\hline
\end{tabular}
\caption{The estimated $1\sigma$ uncertainty of cosmological parameters in $\Lambda \rm{CDM}$ scenario from GC in the linear and MNL regimes. The predictions are compared for EFT (with fixed and marginalized $c_{\rm s}$ and bias parameters) to HF (with fixed and marginalized bias parameters).}
\label{table:sigmalcdm}
\end{table}
\subsection{Results}
\begin{figure}
\centering
\includegraphics[scale=1]{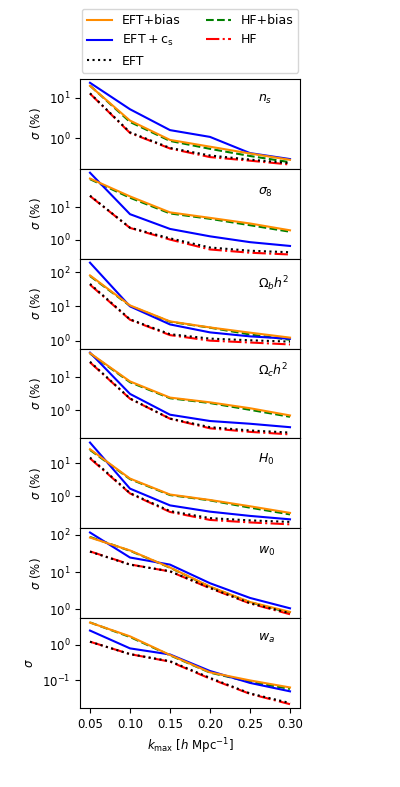} 
    \caption{The estimated $1\sigma$ uncertainties of cosmological parameters in the ($w_0,w_a$)CDM framework for different $k_{\rm max}$ for a Euclid-like survey, with the EFT  and HF  nonlinear recipes, assuming fixed and free $c_{\rm s}$ and bias parameters respectively. All errors are relative except for $w_a$.} 
    \label{fig:sigmak}
\end{figure}

 \textbf{Parameter errors and their $k$-dependence}  Tables~\ref{table:sigmalcdm}, \ref{table:sigmawcdm} and \ref{table:sigmaw0wa} present the estimated parameter errors in the $\Lambda$CDM, $w$CDM and $(w_0,w_a)$CDM frameworks respectively for the GC probe with the specifications proposed 
for Euclid, SKA1 and SKA2 in the two regimes of linear and MNL. 
The EFT results are compared for known and binned bias descriptions, as well as fixed and free  $c_{\rm s}$, while  the HF results are with known and  binned bias parameters.
The presented errors for all parameters  are relative $1\sigma$ errors in percentage, with respect to the parameter fiducial absolute values, expect for $w_a$ whose fiducial is zero and its absolute error is reported.
Figure~\ref{fig:sigmak} illustrates the $k$-dependence of the estimated parameter errors for the $(w_0,w_a)$CDM scenario for a Euclid-like survey. The curves correspond to EFT (with fixed and marginalized $c_{\rm s}$ and bias parameters) and HF (with fixed and marginalized bias parameters). 
\begin{table}
\centering
\setlength\tabcolsep{1.5pt}
\begin{tabular}{l l l l l l l l }
\multicolumn{8}{ c }{$w$CDM} \\ \hline \hline
& &{$n_{\rm s}$}  & $H_0$ & $\Omega _{\rm b}h^2$& $\Omega _{\rm c}h^2$ & $\sigma_8$ & $w$\\
\hline
\textbf{Euclid}\\
\hline
linear  & EFT+$c_{\rm s}$ & $1.6\%$ & $0.5\%$ &$3.0\%$&$0.7\%$&$1.4\%$&$2.6\%$
\\& EFT+bias & $0.9\%$ & $1.0\%$ &$3.2\%$&$2.1\%$&$5.3\%$&$5.4\%$\\
& EFT & $0.6\%$&$0.4\%$&$1.5\%$&$0.5\%$&$0.7\%$&$1.2\%$\\& HF+bias& $0.8\%$ & $1.0\%$ &$3.1\%$&$2.0\%$&$5.0\%$&$5.2\%$\\& HF&$0.6\%$&$0.3\%$&$1.4\%$&$0.5\%$&$0.7\%$&$1.2\%$\\
\hline
MNL & EFT+$c_{\rm s}$ & $0.3\%$ & $0.2\%$ &$1.1\%$&$0.2\%$&$0.4\%$& $0.6\%$
\\& EFT+bias & $0.2\%$ & $0.2\%$ &$1.0\%$&$0.3\%$&$0.9\%$&$0.8\%$\\
& EFT&$0.2\%$&$0.2\%$&$0.9\%$&$0.2\%$&$0.4\%$&$0.5\%$\\& HF+bias & $0.2\%$ & $0.2\%$ &$0.9\%$&$0.3\%$&$1.0\%$&$0.8\%$\\& HF&$0.2\%$&$0.1\%$&$0.8\%$&$0.2\%$&$0.3\%$&$0.5\%$\\
\hline
\textbf{SKA1}\\
\hline
linear  & EFT+$c_{\rm s}$ & $15.8\%$ & $5.3\%$ &$27.4\%$&$8.4\%$&$23.2\%$&$29.2\%$
\\& EFT+bias & $9.1\%$ & $9.9\%$ &$32.8\%$&$21.0\%$&$63.4\%$&$50.9\%$\\
& EFT & $6.3\%$&$3.7\%$&$16.4\%$&$5.2\%$&$10.8\%$&$17.6\%$ \\& HF+bias & $9.0\%$ & $11.1\%$ &$35.3\%$&$24.1\%$&$73.4\%$&$55.0\%$\\& HF&$5.8\%$&$3.2\%$&$14.2\%$&$4.7\%$&$11.5\%$&$19.7\%$\\
\hline
MNL & EFT+$c_{\rm s}$ & $4.0\%$ & $1.1\%$ &$11.4\%$&$3.8\%$&$7.8\%$& $7.3\%$
\\& EFT+bias & $3.1\%$ & $1.1\%$ &$10.2\%$&$4.5\%$&$11.3\%$&$7.2\%$\\
& EFT&$3.0\%$&$0.8\%$&$9.8\%$&$3.2\%$&$6.2\%$&$5.0\%$\\& HF+bias & $3.7\%$ & $1.1\%$ &$7.4\%$&$5.6\%$&$14.1\%$&$7.6\%$\\& HF&$2.7\%$&$0.8\%$&$7.2\%$&$2.9\%$&$5.6\%$&$5.3\%$\\
\hline
\textbf{SKA2}\\
\hline
linear  & EFT+$c_{\rm s}$ & $1.4\%$ & $0.5\%$ &$3.0\%$&$0.8\%$&$1.5\%$&$2.3\%$
\\& EFT+bias & $1.0\%$ & $1.1\%$ &$3.5\%$&$2.3\%$&$5.7\%$&$4.7\%$\\
& EFT & $0.6\%$&$0.4\%$&$1.6\%$&$0.6\%$&$0.8\%$&$1.1\%$ \\& HF+bias & $0.7\%$ & $0.8\%$ &$2.6\%$&$1.6\%$&$4.2\%$&$3.4\%$\\& HF&$0.5\%$&$0.3\%$&$1.2\%$&$0.4\%$&$0.6\%$&$0.8\%$\\
\hline
MNL & EFT+$c_{\rm s}$ & $0.4\%$ & $0.1\%$ &$1.1\%$&$0.3\%$&$0.6\%$& $0.6\%$
\\& EFT+bias & $0.3\%$ & $0.2\%$ &$1.1\%$&$0.5\%$&$1.3\%$&$0.9\%$\\
& EFT&$0.3\%$&$0.1\%$&$1.0\%$&$0.3\%$&$0.5\%$&$0.4\%$\\& HF+bias & $0.3\%$ & $0.2\%$ &$0.8\%$&$0.5\%$&$1.3\%$&$0.8\%$\\& HF&$0.2\%$&$0.1\%$&$0.6\%$&$0.2\%$&$0.3\%$&$0.3\%$\\
\hline
\end{tabular}
\caption{Similar to Table~\ref{table:sigmalcdm} but for a $w$CDM scenario. We used the absolute value of $w$ in the relative error estimation.}
\label{table:sigmawcdm}
\end{table}
\begin{table}
\centering
\setlength\tabcolsep{1.5pt}
\begin{tabular}{l l l l l l l l l}
\multicolumn{9}{ c }{$(w_0,w_a)$CDM} \\\hline \hline
&& {$n_s$}  & $H_0$ & $\Omega _bh^2$& $\Omega _ch^2$ & $\sigma_8$ & $w_0$ & $w_a$\\
\hline
\textbf{Euclid}\\
\hline
linear  & EFT+$c_{\rm s}$ & $1.6\%$ & $0.5\%$ &$3.0\%$&$0.7\%$&$2.2\%$&$15.9\%$&$0.5$
\\& EFT+bias & $0.9\%$ & $1.1\%$ &$3.7\%$&$2.4\%$&$7.0\%$&$13.3\%$&$0.5$\\
& EFT & $0.6\%$&$0.4\%$&$1.5\%$&$0.5\%$&$1.1\%$ &$10.6\%$&$0.2$\\& HF+bias & $0.9\%$ & $1.1\%$ &$3.6\%$&$2.3\%$&$6.5\%$&$13.1\%$&$0.5$\\& HF&$0.6\%$&$0.3\%$&$1.4\%$&$0.5\%$&$1.0\%$&$10.4\%$&$0.3$\\
\hline
MNL & EFT+$c_{\rm s}$ & $0.3\%$ & $0.2\%$ &$1.1\%$&$0.3\%$&$0.6\%$&$1.1\%$&$0.06$
\\& EFT+bias & $0.4\%$ & $0.2\%$ &$1.2\%$&$0.5\%$&$1.3\%$&$1.3\%$&$0.03$\\
& EFT&$0.2\%$&$0.2\%$&$0.9\%$&$0.2\%$&$0.4\%$ &$0.8\%$&$0.01$\\& HF+bias & $0.3\%$ & $0.3\%$ &$1.1\%$&$0.6\%$&$1.8\%$&$0.8\%$&$0.03$\\& HF&$0.2\%$&$0.1\%$&$0.8\%$&$0.2\%$&$0.3\%$&$0.7\%$&$0.02$\\
\hline
\textbf{SKA1}\\
\hline
linear & EFT+$c_{\rm s}$ & $15.8\%$ & $5.6\%$ &$28.3\%$&$8.4\%$&$55.8\%$&$94.6\%$&$6.1$
\\& EFT+bias & $9.4\%$ & $13.0\%$ &$42.7\%$&$25.5\%$&$63.5\%$&$128.5\%$&$5.7$\\
& EFT & $6.8\%$&$4.1\%$&$18.0\%$&$5.2\%$&$27.1\%$&$64.0\%$&$3.5$\\& HF+bias & $9.4\%$ &$13.3\%$&$42.7\%$&$27.8\%$&$73.4\%$&$118.4\%$&$5.0$\\& HF&$5.9\%$&$3.3\%$&$14.8\%$&$4.7\%$&$38.8\%$&$65.0\%$&$4.2$\\
\hline
MNL & EFT+$c_{\rm s}$ & $4.1\%$ & $1.4\%$ &$11.8\%$&$3.8\%$&$10.2\%$& $14.6\%$&$0.8$
\\& EFT+bias & $3.3\%$ & $1.4\%$ &$10.6\%$&$4.6\%$&$13.5\%$&$13.1\%$&$0.7$\\
& EFT&$3.2\%$&$1.2\%$&$10.3\%$&$3.2\%$&$7.9\%$&$12.1\%$&$0.6$\\& HF+bias & $3.7\%$ & $1.8\%$ &$8.8\%$&$5.8\%$&$15.0\%$&$21.5\%$&$1.2$\\& HF&$2.8\%$&$1.3\%$&$8.3\%$&$3.0\%$&$9.7\%$&$17.4\%$&$1.0$\\
\hline
\textbf{SKA2}\\
\hline
linear  & EFT+$c_{\rm s}$ & $1.4\%$ & $0.5\%$ &$3.0\%$&$0.8\%$&$1.5\%$&$7.0\%$&$0.2$
\\& EFT+bias & $1.0\%$ & $1.1\%$ &$3.7\%$&$2.3\%$&$5.8\%$&$8.5\%$&$0.2$\\
& EFT & $0.6\%$&$0.4\%$&$1.7\%$&$0.6\%$&$0.8\%$ &$3.4\%$&$0.1$\\& HF+bias & $0.7\%$ & $0.8\%$ &$2.7\%$&$1.7\%$&$5.0\%$&$6.0\%$&$0.2$\\& HF&$0.5\%$&$0.3\%$&$1.3\%$&$0.4\%$&$1.0\%$&$4.5\%$&$0.2$\\
\hline
MNL & EFT+$c_{\rm s}$ & $0.4\%$ & $0.2\%$ &$1.3\%$&$0.3\%$&$0.6\%$&$1.3\%$&$0.04$
\\& EFT+bias & $0.3\%$ & $0.2\%$ &$1.2\%$&$0.5\%$&$1.3\%$&$1.3\%$&$0.03$\\
& EFT&$0.3\%$&$0.2\%$&$1.1\%$&$0.3\%$&$0.5\%$ &$1.0\%$&$0.01$\\& HF+bias & $0.3\%$ & $0.2\%$ &$0.9\%$&$0.5\%$&$1.3\%$&$1.2\%$&$0.03$\\& HF&$0.2\%$&$0.1\%$&$0.8\%$&$0.2\%$&$0.3\%$&$1.0\%$&$0.03$\\
\hline
\end{tabular}
\caption{ Similar to Table~\ref{table:sigmawcdm} but for a $(w_0,w_a)$CDM scenario.  The errors on $w_a$ are absolute.}
\label{table:sigmaw0wa}
\end{table}
As we go to the MNL regime, corresponding to higher $k$'s, the uncertainty in the cosmological parameters are significantly reduced indicating the large  amount of information available in these small scales.
In the three cosmic scenarios, and with the bias and $c_{\rm s}$ fixed, the uncertainties are very close for HF and EFT in the linear regime as expected.
The tiny mismatch for some parameters, with HF slightly underestimating the error, is not surprising since nonlinearities are not fully negligible at $k\sim 0.15~h~{\rm Mpc}^{-1}$. 
The inclusion of larger wavenumbers significantly improves the measurements.
The estimated errors are still comparable for the two MNL recipes, with the HF-based analysis underestimating some of the uncertainties. 
On the other hand, if the tight prior on the bias and $c_{\rm s}$ are relaxed and they are allowed to vary, there is notable boost in the uncertainties which varies for different parameters. It is interesting to note that this boost is significant for some parameters even in the linear regime. This is induced by the correlation between the $c_{\rm s}$  and some of the cosmological parameters which enhances the errors  due to lack of information in the linear regime. In the MNL regime, with  more data available, the correlations partially break and the errors decrease.  \\

\textbf{Correlation matrix} The inclusion of larger wavenumbers  affects the parameter distribution in the parameter space and  therefore their measurements not only through the reduced errors but also through the induced correlation between different parameters. We explore this by the correlation matrix $P_{ij}=C_{ij}\slash \sqrt{C_{ii}C_{jj}}$, where the off-diagonal terms illustrate the degree of correlations. In the case of fully uncorrelated parameters, $\boldsymbol{P}$ would be the identity matrix.

\begin{figure*}
\includegraphics[width=\textwidth,height=\textheight,keepaspectratio]{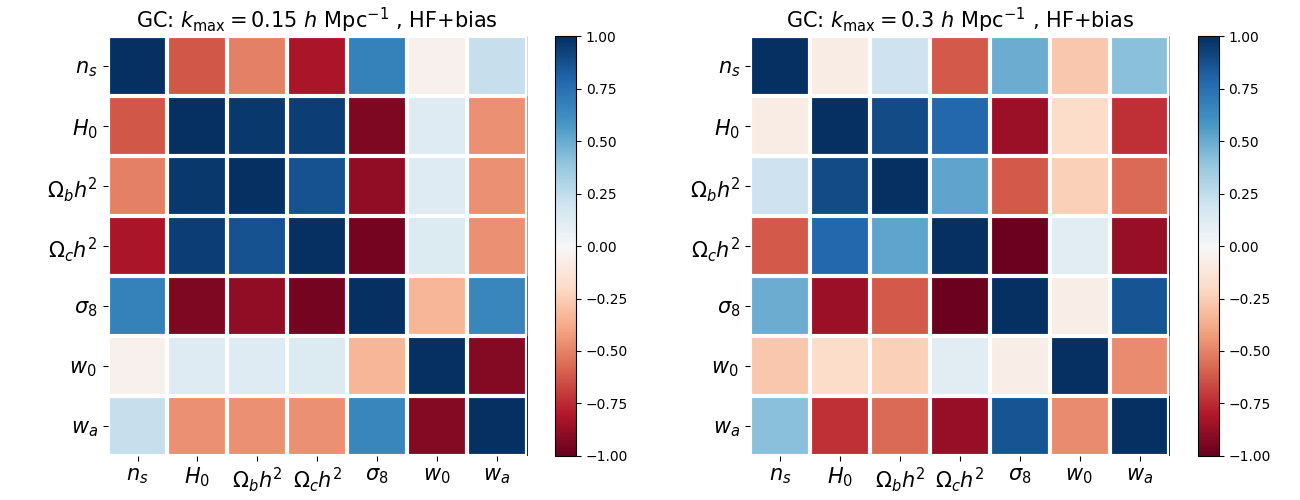}
\caption{Correlation matrix $\boldsymbol{P}$ for the parameters in the $(w_0,w_a)$CDM scenario with the HF nonlinear prescription, in the linear with $k_{\rm max}=0.15~h~\rm{Mpc}^{-1}$ (left) and MNL  with $k_{\rm max}=0.3~h~\rm{Mpc}^{-1}$ (right) regimes.}
\label{fig3}
\end{figure*}

\begin{figure*}
\includegraphics[width=\textwidth,height=\textheight,keepaspectratio]{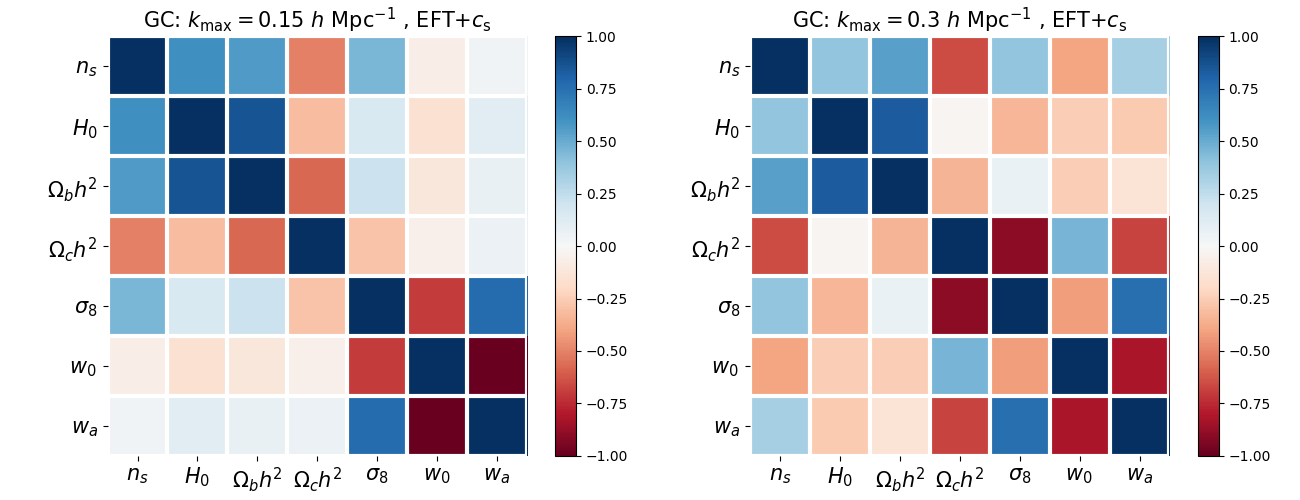}
\caption{Similar to Figure~\ref{fig3}  but with EFT nonlinear prescription.}
\label{fig4}
\end{figure*}

Figures~\ref{fig3} and \ref{fig4} compare how smaller scales, up to $k_{\rm max} =0.3~h~{\rm Mpc}^{-1}$, would affect the parameter correlations for HF and EFT, marginalized over $c_{\rm s}$ and bias parameters respectively, compared to the linear case with $k_{\rm max} =0.15~ h~{\rm Mpc}^{-1}$. 
As is evident from the figures, the parameter correlations decrease in most cases as more information becomes available with larger wavenumbers. 
However, it is also interesting to note that this correlation decrease is not a general rule. 
With more informative data in hand, the errors on cosmological parameters diminish. However, the degeneracy could increase for some parameters. 
 This could be explained as the following. 
 For these parameters, the large scale imprints are distinguishable, although  not tightly constrained.
 The small scales, on the other hand, yield a lot more, yet not fully distinct, information about the parameters of our interest.  
Therefor, for these cases, as the errors decrease, the parameter degeneracy increases.\\ 

\textbf{Figure of merit} The overall ability of a particular experiment in constraining the  volume spanned by the parameters in the parameter space can be quantitatively described  through a figure of merit ($\rm FoM$). Here we define the $\rm FoM$ as \citep{albrecht06}
\begin{equation}
\rm FoM=-\frac{1}{2}\ln(\det \boldsymbol{C}).
\end{equation}
The stronger the constraints on cosmological parameters, the smaller the corresponding volume in the parameter space, and therefor the  determinant of $\bf C$,  which leads to bigger FoM. Here we compute FoM as a function $k_{\rm max}$ with HF and EFT as our theoretical nonlinear prescriptions.

\begin{figure*}
    \centering
    \includegraphics[scale=0.57]{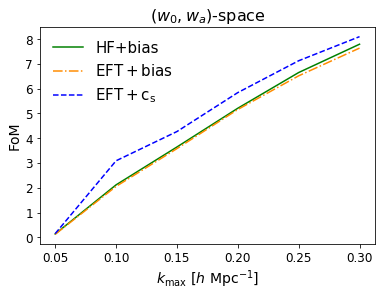}    
    \includegraphics[scale=0.57]{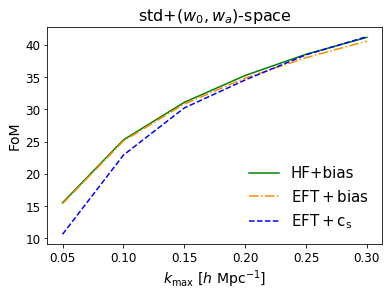}    
    \caption{The $\rm FoM$ in the $(w_0, w_a)$CDM cosmological model, compared for the HF and EFT nonlinear prescriptions.
    Left: The $\rm FoM$ for the $(w_0,w_a)$ parameter space, marginalized over the cosmological and nuisance parameters.
      Right: The $\rm FoM$ for the full cosmological parameters, consisting of $(w_0,w_a)$ and the standard cosmic parameters, marginalized over the nuisance parameters.} 
\label{fig5}
\end{figure*}

Figure~\ref{fig5} illustrates the $\rm FoM$ in the $(w_0, w_a)$CDM scenario with different assumptions for the parameter space. The left panel presents the improvement of the FOM in the overall measurement of the equation of state of dark energy as more modes are included in the analysis. The results are marginalized over the rest of cosmological and nuisance parameters. 
The right panel presents similar results but for the full space of cosmological parameters. 
\section{Discussion}\label{sec:conclude}
The upcoming large scale surveys will deliver huge amount of information in the mildly nonlinear regime. The analysis of this high precision data requires careful modeling of the nonlinear regime for reliable cosmological inferences. 
In this work we investigated the bias induced in the parameters when estimated  on the basis of the popular Halofit nonlinear model. The bias is most significant for $\sigma_8$ and $\Omega_{\rm b}h^2$ with $\sim (4-5)\sigma$, depending on the cosmological scenario, from the galaxy clustering probe of a future Euclid-like survey when modes up to $k_{\rm max}=0.3~h~{\rm Mpc}^{-1}$ are included. 
This clearly demonstrates how the few percent inaccuracy in the nonlinear model translates into a few $\sigma$ bias in the parameter measurements for the near future high precision large scale surveys. 

The above analysis was performed with the only free parameters being the cosmologically interesting parameters. 
A more realistic analysis, however, often includes marginalization over many nuisance parameters. The inclusion of nuisance parameters could increase the estimated uncertainties and reduce the parameter bias. 
We explored the sensitivity of the forecasted errors, as a function of the maximum wavenumber included in the analysis, to various experimental setups as well as different nonlinear prescriptions and nuisance parameters.
In particular we used the EFTofLSS as the main nonlinear model with its so-called sound speeds as the nuisance parameters. We compared the results with those based on Halofit, with the redshift dependence of the galaxy bias assumed unknown and blindly parametrized to be measured by the data itself. 
The estimated parameter uncertainties from the two recipes are found comparable up to $k_{\rm max}=0.3~h~\rm{Mpc}^{-1}$. This is evident from the HF and EFT curves of Figure~\ref{fig5}, where the same sets of parameters are included in the analysis with the two nonlinear recipes. 
It should however be noted that the comparability of the estimated Fisher-based errors simply indicates the similar power of the two approaches in extracting cosmic information from the power spectrum. 
Obviously, this should not  be taken as implying similarity in the predictions of the nonlinear recipes when applied to data, as was seen from biased best-fit measurements of the HF-based analysis. 

Relaxing the prior assumptions on the EFT sound speeds and galaxy bias would increase the estimated uncertainty and reduce the parameter biases discussed in Section~\ref{sec:bias}. This is most significant for $\sigma_8$, $\Omega_{\rm c}h^2$ and $w_a$ where the estimated errors boost significantly with marginalization over bias parameters at $k_{\rm max}=0.3~h~\rm{Mpc}^{-1}$.
The nuisance parameters of EFT sound speed are also found to increase the errors (e.g., see the case for $w_a$), but often to a lesser extent compared to the galaxy bias. 
A full analysis would include marginalization over both of these nuisance parameter sets. 
The high impact of these nuisance parameters on the measurement of the cosmologically interesting parameters clearly indicates the need for their appropriate modeling. Tight priors tighten the cosmological constraints and make  inferences more concluding. However, if the priors are biased, the final inferences may  markedly deviate form the true underlying scenario. 
A semi-blind parametrization of these nuisance parameters, on the other hand, could significantly boost the errors and hence reduce the parameter bias which could arise due to inappropriate choice of nonlinear prescription.
This, in turn, renders the difference between the nonlinear recipes practically less effective.

\bibliography{nlgc}
\bibliographystyle{apj}

\end{document}